\documentclass[aip,jcp,
	  sd,%
	   amsmath,amssymb,
	    reprint,%
    ]{revtex4-1}

\usepackage{float}              
\usepackage{supertabular}
\usepackage{epsfig}
\usepackage{rotating}
\usepackage{exscale}            
\usepackage{graphicx}           
\usepackage{epstopdf}
\usepackage[rm,hang]{caption}   
\usepackage{amsfonts}           
\usepackage{ifthen}
\usepackage{xspace}
\usepackage{booktabs}
\usepackage[footnotesize]{subfigure}
\usepackage[dvips]{psfrag}
\usepackage{color,colortbl}
\usepackage{hyperref}
\usepackage{threeparttable}
\usepackage{dcolumn}% Align table columns on decimal point
\usepackage{bm}

\newcommand{\fref}[1]{Figure~\ref{#1}}

\newcommand{\eref}[1]{Equation~\eqref{#1}}

\begin{document}
\preprint{AIP/123-QED}

\title{\Large \bf 
Exploring High Dimensional Free Energy Landscapes: Temperature Accelerated Sliced Sampling
}

\author{Shalini Awasthi }
 \affiliation{Department of Chemistry, Indian Institute of Technology, Kanpur, 208016, India }%
 
\author{ Nisanth N. Nair}
 \homepage{Corresponding Author: nnair@iitk.ac.in}
 \affiliation{Department of Chemistry, Indian Institute of Technology, Kanpur, 208016, India }%

 \date{\today}% It is always \today, today,

 \begin{abstract}
Biased sampling of collective variables is widely used to accelerate rare events in molecular simulations and to explore   free energy surfaces. 
However, computational efficiency of these methods decreases with increasing number of collective variables,
which severely limits the predictive power of the enhanced sampling approaches. 
Here we propose a method called Temperature Accelerated Sliced Sampling (TASS) that combines temperature accelerated molecular dynamics with umbrella sampling and metadynamics  to sample the collective variable space in an efficient manner. 
The presented method can sample a large number of collective variables and is advantageous for controlled exploration of broad and unbound free energy basins.
%
%and permits to add or remove collective variables rom enhanced sampling at various umbrella windows. 
%
TASS is also shown to achieve quick free energy convergence and is practically usable with {\it ab initio} molecular dynamics techniques.
 \end{abstract}

% \pacs{Valid PACS appear here}% PACS, the Physics and Astronomy
 
\maketitle

\section{Introduction}
In a canonical ensemble molecular dynamics (MD) simulation, configurations are sampled with the probability 
\begin{eqnarray}
P(\mathbf R) = \frac{e^{-\beta U(\mathbf R) } }{Z} \nonumber
\end{eqnarray}
where $\mathbf R$ is the configuration of a molecular system with $N$ number of atoms, $\beta= 1/k_{\rm B} T$ with Boltzmann constant $k_{\rm B}$ and temperature $T$. Here $U$ is the potential energy, and $Z$ is the configurational partition function. Let  the order parameter be $\zeta(\mathbf R)$, then the probability along $\zeta$ is given by
\begin{eqnarray}
P(\zeta^\prime) = \frac{1}{Z} \int d \mathbf R \, \delta \left ( \zeta(\mathbf R) - \zeta^\prime \right ) e^{-\beta U(\mathbf R) } \enspace . \nonumber
\end{eqnarray}
The Helmholtz free energy along $\zeta$ can then be computed as
\begin{eqnarray}
F(\zeta) = - \frac{1}{\beta} \ln P(\zeta) + f \enspace , \nonumber
\end{eqnarray}
where $f$ is some constant. 
%$F(\zeta)$ provides information on the spontaneity of the process, and in principle, 
$F(\zeta)$ could be directly obtained from the probability distribution of $\zeta$ computed from a canonical ensemble MD simulation, 
provided  a proper sampling of $\zeta$ is achieved.~\cite{Tuckerman:Book,Gabriel:Book,Vanden:2009,Christ:2010,Bussi:Entropy} 
%I added reviews related for the  computation free energy. TODO3
%TODO-SA: can you check if there is some review that an be cited here for rare-event in general.

Often it is more convenient to assume that $\zeta$ is a linear combination of a few collective variables $\{ S_\alpha(\mathbf R)  \}$. In practice, probability distribution $P(\mathbf S)$ for the set of selected collective variables is constructed as,
\begin{eqnarray}
P(\mathbf S^\prime) = \frac{1}{Z} \int d \mathbf R \, e^{-\beta U(\mathbf R) } \prod_{\alpha} \delta \left ( S_\alpha(\mathbf R) -  S_\alpha^\prime \right ) \enspace , \nonumber
\end{eqnarray}
thus
\begin{eqnarray}
F(\mathbf S) = - \frac{1}{\beta} \ln P(\mathbf S) + f \nonumber
\end{eqnarray}
and the minimum energy pathway can be traced on the multi-dimensional surface $F(\mathbf S)$. 
This assumes that we have the knowledge of $\mathbf S$ for describing the process of our interest. 
The current work presumes that the set of collective variables $\left \{ S_\alpha \right \}$ to describe and to sample the distribution  is known, however, the number of collective variables is large. 
%
%Recently, there has been several efforts to obtain information about the vital coordinates to describe a process adequately.{\tt [cite Pratyush's recent papers on this, Ensing's PRL etc.]}
Although, the number of coordinates to describe a process is often small in number,\cite{Miller:2014,Pratyush:2016} 
 several other orthogonal coordinates have to be enhanced-sampled for a quick convergence in probability
 distribution along the reactive coordinates and thus the free energy estimates. 
%TODO
%"10.1146/annurev-physchem-040412-110117"
%also cite annurev of omarvalsson/pratyush/parrinello. (recent one)

Timescale at which a barrier crossing event takes place on a potential energy landscape during a canonical ensemble simulation is $\propto  e^{\beta U(\mathbf R) }$. 
Due to the limitation of small time steps in MD simulations, the simulation time to 
observe such processes becomes very large and computationally unfeasible for many interesting processes with free 
energy barrier $\Delta F^\ddagger >> \beta^{-1}$. 
One of the ways in which this timescale bottleneck can be overcome is by modifying the Boltzmann weight 
%:  a) enhancing the sampling with $\beta^\prime$, where $\beta^\prime <<  \beta$; b) enhancing the sampling 
through altering $U(\mathbf R)$ as $U(\mathbf R) + U^{\rm bias}(\mathbf S)$ where $U^{\rm bias}(\mathbf S)$ is the bias potential. 
Metadynamics\cite{Laio:PNAS:02,Iannuzzi:03,mtd:rev:11,Luigi:12,Gervasio:08} (MTD) and Umbrella Sampling (US)~\cite{us:orig,Kastner:11} are two such popular biased sampling methods, among 
several others~\cite{Huber:1994,Grubmuller:PRE:95,Wang:2001,Drave:2001,Hansmann:2002,Hu:IS,Chipot:2015,Gao:IS}. %TODO7: cite other biased sampling papers here - may be look at the tuckerman's paper or recent reviews for all the references.

In MTD, a time dependent bias potential, $U^{\rm bias}\equiv V^{\rm b}(\mathbf S,t)$, is constructed by summing the Gaussian potentials deposited discretely along the
trajectory $\mathbf S(t)$:
%increases the sampling in CV space:
\begin{eqnarray}
%\label{eqn:mtd:2}
V^{\rm b}(\mathbf S,t) = \sum_{\tau < t} w_{\tau} \exp \left [ - \frac{ \left \{ \mathbf S - \mathbf S(\tau) \right \}^2 }{2 (\delta s)^2 } \right ],
\nonumber
\end{eqnarray}
In the Well Tempered (WT--MTD)~\cite{mtd:well:08} variant of MTD,
\begin{eqnarray}
%\label{eqn:mtd:3}
w_{\tau}=w_0 \exp \left [ - \frac{V^{\rm b}(\mathbf S, t)}{ k_{\rm B} \, \Delta T }\right ] \nonumber
\end{eqnarray}
where $w_0$ is the initial Gaussian height and $\Delta T$ is a parameter. Free energy estimate can be obtained as~\cite{Voth:14}
\begin{eqnarray}
%\label{eqn:mtd:4}
F(\mathbf S) = - \gamma \lim_{t \rightarrow \infty}  V^{\rm b}(\mathbf S,t) + f \nonumber 
\end{eqnarray}
where 
\begin{eqnarray}
\label{gamma:wtmtd}
 \gamma = {(T + \Delta T)}/{\Delta T}
 \end{eqnarray}
  and $f$ is some constant.

The main advantage of MTD is that it is capable of sampling the $\mathbf S$ space in a self--guided manner, and thus the method can explore
unprecedented minima and reaction pathways on 
high--dimensional free energy landscapes.\cite{Gervasio:08,mtd:rev:11,Pratyush:2016}
 Nowadays, MTD is used in exploring free energy landscapes up to three collective variables. 
 The total computational time required to explore the free energy landscape depends exponentially on the 
 number of collective variables.
% %
% Although, the number of crucial collective coordinates may be small in number for describing the reaction coordinate
% or an order parameter, there could exist large number of 
% orthogonal collective coordinates, which also needs to be sampled efficiently for proper free energy estimates and exploration.
  %
 In order to increase the efficiency of sampling large number of coordinates, parallel tempering MTD~\cite{Bussi:2006}, bias--exchange MTD\cite{be:mtd:1,be:mtd:2}, replica exchange with collective variable tempering~\cite{Bussi:2015}, parallel bias MTD~\cite{Pfaendtner:2015}, and variational MTD~\cite{Omar:2016} methods have been proposed. %TODO2 more methods can be added; and review of Pratyush could be also cited if you find this is mentioned there.

In US simulations, a time independent harmonic restraint bias potential, $U^{\rm bias} \equiv W_h^{\rm b}$, is applied at chosen discrete values of $\mathbf S$, given by
\begin{eqnarray}
\label{eqn:us:1}
W^{\rm b}_h(\mathbf S) = \frac{1}{2} \kappa_h \left ( \mathbf S - \mathbf S_{h} \right )^2, \enspace \enspace h=1,\cdots, M  
 \end{eqnarray}
 where $\mathbf S_h$ is the position of the umbrella window $h$. To obtain $F(\mathbf S)$, the distribution of $\mathbf S$ from $M$ windows are reweighted and stitched together by the weighted histogram analysis (WHAM) method.\cite{wham:1,wham:2} 
The sampling of the collective variables are determined by the span of the windows, and thus 
US allows to achieve a controlled sampling of collective variable space.
Like in MTD, the computational cost increases with the number of dimensions and most of the applications using this
technique have been limited to one or two collective variables only.

Another way to accelerate the sampling of collective variables is by modifying the Boltzmann factor using $ \tilde \beta << \beta$,  where $\tilde \beta$ corresponds to the temperature $\tilde T$, which is much greater than the system temperature $T$.
This is achieved in Temperature Accelerated Molecular Dynamics (TAMD)\cite{TAMD:1,Tuckerman:2008,Tuckerman:Book} approach by defining %TODO cite van de Eijenden, then tuckerman, tuckerman-book 
an extended system where a set of auxiliary variables $\left \{ s_\alpha \right \}$ 
is introduced that couple with $\left \{ S_\alpha \right \}$ through a harmonic potential.
Further, $\left \{ s_\alpha \right \}$ is thermostated to $\tilde \beta$, while the physical system is thermostated to $\beta$, 
and the free energy at $\beta$ can be computed as,\cite{Tuckerman:2008,Tuckerman:Book} 
\begin{eqnarray}
\label{e:f:tamd}
	F(\mathbf s)=-\frac {1}{\tilde \beta}\ln  \tilde P(\mathbf s) + f
\end{eqnarray}
where $\tilde P(\mathbf s)$ is the probability distribution of $\left \{ s_\alpha \right \}$ computed at $\tilde \beta$.
Tuckermann and co--workers~\cite{Tuckerman:2012} have integrated TAMD with biased sampling approach to 
improve its efficiency and further extended this approach to build a free energy minimization procedure to locate 
saddle points and minimum energy pathways on complex free energy landscapes.~\cite{Tuckerman:2015} 
It may be noted that in their ``heating and flooding'' approach, both 
temperature acceleration and the bias potentials are applied simultaneously to all the collective variables.

We have recently introduced a method called Well--Sliced MTD (WS--MTD)~\cite{shalini:2016} to overcome the limitation of metadynamics in sampling broad and unbound free energy basins which are encountered often in the case of A+B type of chemical reactions, drug binding, protein folding etc.
In this technique, we have combined US and MTD to sample orthogonal collective variables simultaneously.
US allows to achieve controlled sampling of collective variables, while MTD allows to sample orthogonal variables in a
self--guided manner.
However, the efficiency of this approach also decreases with increasing number of collective variables.

In the current work, we introduce a technique called Temperature Accelerated Sliced Sampling (TASS), which extends the WS--MTD approach to explore free energy landscape with large number of collective variables.
The efficiency is improved by introducing temperature acceleration of collective variables in the spirit of TAMD.
The method could be considered as an improvement to MTD and TAMD approaches to sample broad and unbound surfaces in an efficient
manner.
Furthermore, this method may also be looked at as an extension to the US for incorporating large number of orthogonal coordinates.
At first, we will discuss the theory behind the TASS approach, and then demonstrate its efficiency for the following four problems:
(a) exploring a  three  dimensional potential model; 
(b) computing the free energy landscape in the space of four backbone torsions for alanine tripeptide {\em in vacuo} using the AMBER force--field; 
(c) modeling cyclization reaction of butadiene using {\em ab initio} Car--Parrinello MD by sampling 
three collective variables; 
(d) computing the free energy barrier for the hydrolysis reaction of an enzyme--drug complex by sampling four collective variables 
in a density functional theory (DFT) based QM/MM MD simulation.

\section{Theory}
\label{s:methods}
In the TASS approach, we use the Hamiltonian
%\begin{widetext}
\begin{eqnarray}
\label{eqn:ham}
H_h (\mathbf R, \mathbf P, \mathbf s,\mathbf p) &=&   H^0(\mathbf R,\mathbf P) \nonumber \\
                              & + & \sum_{\alpha=1}^{n}\left [ \frac{ p_\alpha^2}{2 \mu_\alpha} +	   \frac {k_\alpha} {2} \left (S_\alpha({\bf R})-s_\alpha \right )^2 \right ]   \nonumber \\
	              & & + W^{\rm b}_h(s_1) + V^{\rm b}_h(s_2,t)    \nonumber \\ 
	               & &   + \mathrm{bath}(\mathbf P; T)  + \mathrm{bath}(\mathbf p; \tilde T) \enspace ,  
			      \end{eqnarray}
%\end{widetext}
where $h=1,\cdots, M$  and $n \geqslant 2$. 
Here $H^0$ is the system Hamiltonian, 
$\mathbf R$ and $\mathbf P$ are the set of all atomic positions and momenta, and 
$\mathbf S(\mathbf R)$ is the set of $n$  collective variables. 
Importantly, $n$ number of auxiliary variables $\left \{ s_\alpha \right \}$ with masses $\{ \mu_\alpha \}$ and momenta $\{ p_\alpha \}$ 
are introduced that couple to the collective variables $\left \{ S_\alpha \right \}$ 
by a harmonic potential with coupling constants $\{ k_\alpha \}$. 
Along $s_1$ and $s_2$, umbrella and metadynamics bias potentials $W^{\rm b}_h(s_1)$ and $V^{\rm b}_h(s_2,t)$ 
are added, respectively.
The atomic system is coupled to a thermal bath at temperature $T$ and auxiliary variables are coupled to a thermostat at temperature $\tilde T$.
 Also, $\{ k_\alpha \}$ and $\{ \mu_\alpha \}$ values are chosen such that  the dynamics of $\{ s_\alpha \}$ is close to $\{ S_\alpha \}$ and they are 
 adiabatically decoupled as done in a regular TAMD simulation.~\cite{TAMD:1,Tuckerman:2008,Tuckerman:Book}
Our aim is to construct the free energy landscape $F(\mathbf s)$ at temperature $T$. 
%which should be nearly the same as $F(\mathbf S)$ at temperature $T$.
%

Hamiltonian in \eref{eqn:ham}, allows one to sample the collective variables space by a combination of US, MTD, and  TAMD.
Especially, the temperature accelerated sampling in the spirit of TAMD allows one to choose larger number of collective variables compared to other biased
sampling techniques.
In \eref{eqn:ham}, only one collective variable is biased using US and MTD, however, more number of variables could be biased in a straightforward manner.

Following the reweighting equations for WS--MTD as used in our previous work~\cite{shalini:2016},  %TODO 
we first reweight the metadynamics bias potential~\cite{Bonomi:09,Tiwary:14} as 
%\begin{widetext}
\begin{eqnarray}
\label{e:p:unb}
\tilde P_h (\mathbf s^\prime) =  \frac{ \int d \tau \, A_h(\tau) 
%\exp[  \tilde \beta \left \{ V_h^{\rm b} (s_2(\tau),\tau) - c_h(\tau) \right \} ] 
\prod_\alpha^n \enspace \delta(s_\alpha(\tau) - s_\alpha^\prime) 
}
                             {\int d\tau \, A_h(\tau) 
                             %\exp[ \tilde \beta \left \{ V_h^{\rm b}(s_2(\tau),\tau) - c_h(\tau) \right \} ]  
                             } \enspace ,
\end{eqnarray}
%\end{widetext}
where 
\[ A_h(\tau) = \exp[ \tilde \beta \left \{ V_h^{\rm b}(s_2(\tau),\tau) - c_h(\tau) \right \} ] \]
with
\begin{eqnarray}
%\label{e:ct}
c(t) = \frac{1}{\tilde \beta} \ln \left [ \frac{ \int d s_2 \exp [  \tilde \beta \gamma V^{\rm b}(s_2,t) ] } 
{ \int ds_2 \exp [  \tilde \beta \left ( \gamma - 1 \right ) V^{\rm b}(s_2,t) \} ] } \right ] \enspace , \nonumber
\end{eqnarray}
and the constant $\gamma$ is given by \eref{gamma:wtmtd}.
In the subsequent step, we reweight $\tilde P_h (\mathbf s)$ for the umbrella bias potential and combine distributions of all umbrella windows using
the standard WHAM approach. 
In this procedure, the reweighted distribution $\tilde P(\mathbf s)$ is obtained from $M$ number of $\tilde P_h(\mathbf s)$ using a self-consistent
approach using
\begin{eqnarray}
%\label{e:wham}
\tilde P(\mathbf s) = \frac{\sum_{h=1}^{M}  n_h \tilde P_{h}(\mathbf s) }
                             {\sum_{h=1}^{M} n_h  \exp[ \tilde \beta  f_h]   \exp[ - \tilde \beta  W_h(s_1) ] }  \enspace , \nonumber
\end{eqnarray}
with
\begin{eqnarray}
%\label{eqn:fk:u}
\exp[{- \tilde \beta  f_h} ] = \int ds_1  \exp[ {- \tilde \beta W^{\rm b}_h(s_1) }]  \tilde P(\mathbf s) \enspace , \nonumber
\end{eqnarray}
and $W^{\rm b}_h$ is given by \eref{eqn:us:1}.
Here $n_h$ is the number of configurations sampled in the $h^{\rm th}$ window of the umbrella potential.
If the collective variables and the auxiliary variables are adiabatically separated, 
the distribution $\tilde P(\mathbf s)$ at higher temperature $\tilde T$ is related to $P(\mathbf s)$
at temperature $T$ as~\cite{Tuckerman:2008} 
\begin{eqnarray}
%\label{e:prob}
P(\mathbf s) \propto \tilde P(\mathbf s)^{(\beta/\tilde \beta )}  \enspace. \nonumber
\end{eqnarray}
Then, free energy surface at temperature $T$ can be obtained using \eref{e:f:tamd}.

\section{Results and Discussion}
\label{s:res}

\subsection{Three Dimensional Model System}
\label{s:res:model}

For testing the method, we considered a three--dimensional model system that has four minima: 
%\begin{widetext}
\begin{eqnarray}
U(x,y,z) &=& \sum_{i=1}^{3} U_i^o  \exp \left (-w_i \left [(x-x_i^o)^2 + \right . \right . \nonumber \\ 
& &  \left . \left . b_i(y-y_i^o)^2 + c_i(z-z_i^o)^2 \right ] \right) \nonumber
\end{eqnarray}
%\end{widetext}

\begin{figure*}[t]
\begin{center}%
\includegraphics[width=0.7\textwidth]{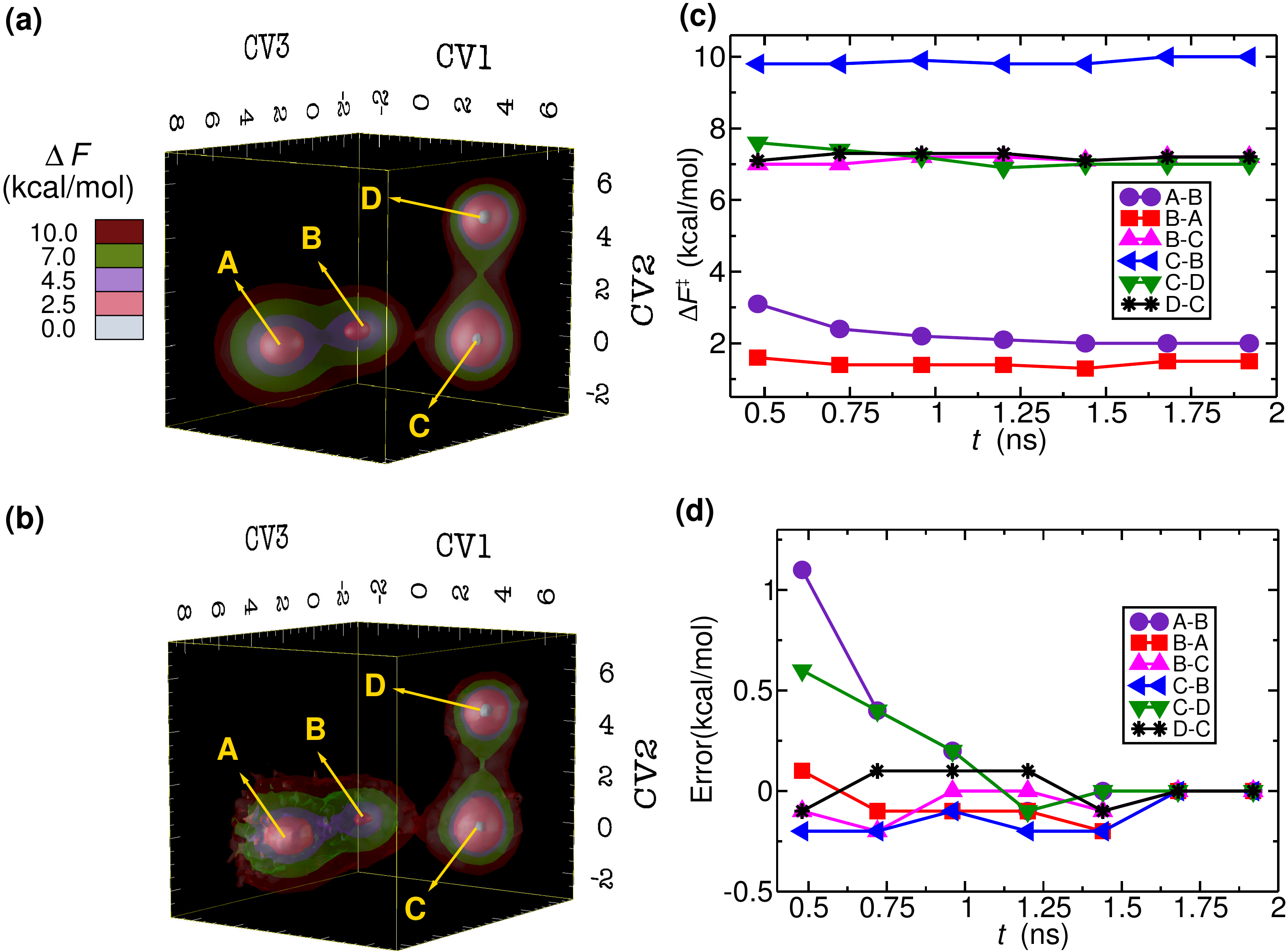}
\end{center}%
\caption[]{%
\label{model_FES}%
(a) Three--dimensional model potential with four minima used to test the TASS method, visualized as contour surfaces; 
(b) Free energy surface reconstructed using the TASS method; 
(c) Free energy barriers computed from TASS as a function of simulation time per umbrella window; 
(d) Exact error in the free energy barrier estimates as a function of simulation time per umbrella window.}
\end{figure*}

Parameters for the potential are given in Table~SI1, and the plot of $U(x,y,z)$ is shown  
in \fref{model_FES}a.
The four minima are labeled as {\bf A}, {\bf B}, {\bf C} and {\bf D}, and the barriers in this potential energy landscape are  tabulated in 
Table~SI2.
%\tref{table:model:parameters}. %TODO
% {\bf A}$\rightarrow${\bf B} and {\bf B}$\rightarrow${\bf A} are 2.0~kcal~mol$^{-1}$ and 1.5~kcal~mol$^{-1}$, respectively, while barriers for {\bf B}$\rightarrow${\bf C} and {\bf C}$\rightarrow${\bf B} are 7.2~kcal~mol$^{-1}$ and 10.0~kcal~mol$^{-1}$, respectively.
%
%Barriers for {\bf C}$\rightarrow${\bf D} and {\bf D}$\rightarrow${\bf C} are 7.0~kcal~mol$^{-1}$ and 7.2~kcal~mol$^{-1}$, respectively.
%
Mass of the system was taken as $1.0$~a.m.u. and MD time step was chosen as $0.24$~fs.

In the TASS simulation, $x$, $y$, and $z$ coordinates were chosen as collective variables; i.e. $S_1\equiv x $, $S_2 \equiv y$, and $S_3 \equiv z$.
Masses of auxiliary variables $\{s_\alpha \}$ were taken as $40.0$~a.m.u.  
and values of $k_\alpha$ were taken as $3.14\times 10^3$~kcal~mol$^{-1}$~Bohr$^{-2}$.

Temperature acceleration was then invoked along all the  auxiliary variables $\{ s_\alpha \}$.
The system temperature was set to 300~K, while that of the auxiliary variables was set to $600$~K.
Temperature of the system and that of the auxiliary variables were maintained using two separate Langevin thermostats with frictional coefficients $0.02$~fs$^{-1}$ and $0.04$~fs$^{-1}$, 
respectively.
Auxiliary variables $s_1$ and $s_2$ were (arbitrarily) chosen for applying US and MTD biases, respectively. 
Umbrella potentials were placed along $s_1$ from $-0.5$~Bohr to $6.5$~Bohr at intervals of $0.5$~Bohr.
Initial structure for any given umbrella window was generated by setting the $s1$ coordinate to that
corresponding to the equilibrium value of the umbrella window, while the other
coordinates were having the same values as in the minimum {\bf A}.
Restraining potential $\kappa_h$ used for all the umbrella potentials was 
%$3.14\times 10^1$~kcal~mol$^{-1}$~Bohr$^{-2}$.
$31.4 $~kcal~mol$^{-1}$~Bohr$^{-2}$. 
The initial Gaussian height ($w_0$) was set to $0.6$~kcal~mol$^{-1}$ and the Gaussian width parameter $\delta s$ was $0.5$~Bohr.
The parameter $\Delta T$  was taken as $1200$~K.
MTD bias potential was updated every $200$~MD steps.
%

%A total of 15 umbrella windows were placed along the $s_1$ coordinate and
%MTD simulations were carried out along $s_2$.
%
The convergence of free energy barriers
as a function of simulation length (per umbrella window) is shown in Table~SI2 and \fref{model_FES}c.
From \fref{model_FES}d, it is clear that the free energy estimates converge to the exact result with increase in simulation
time.
The converged free energy surface is also plotted in \fref{model_FES}b.
Positions of these minima and the topology of the potential energy surface are correctly reproduced in the reconstructed free energy surface. 

These results show that a free energy surface with multiple minima and complex topology can be
efficiently explored by the TASS method.
Moreover, the free energy estimates systematically converge to the exact results. 
\subsection{Alanine Tripeptide}
\label{s:res:alatrip}

\begin{figure*}[]
\begin{center}%
\includegraphics[width=0.7\textwidth]{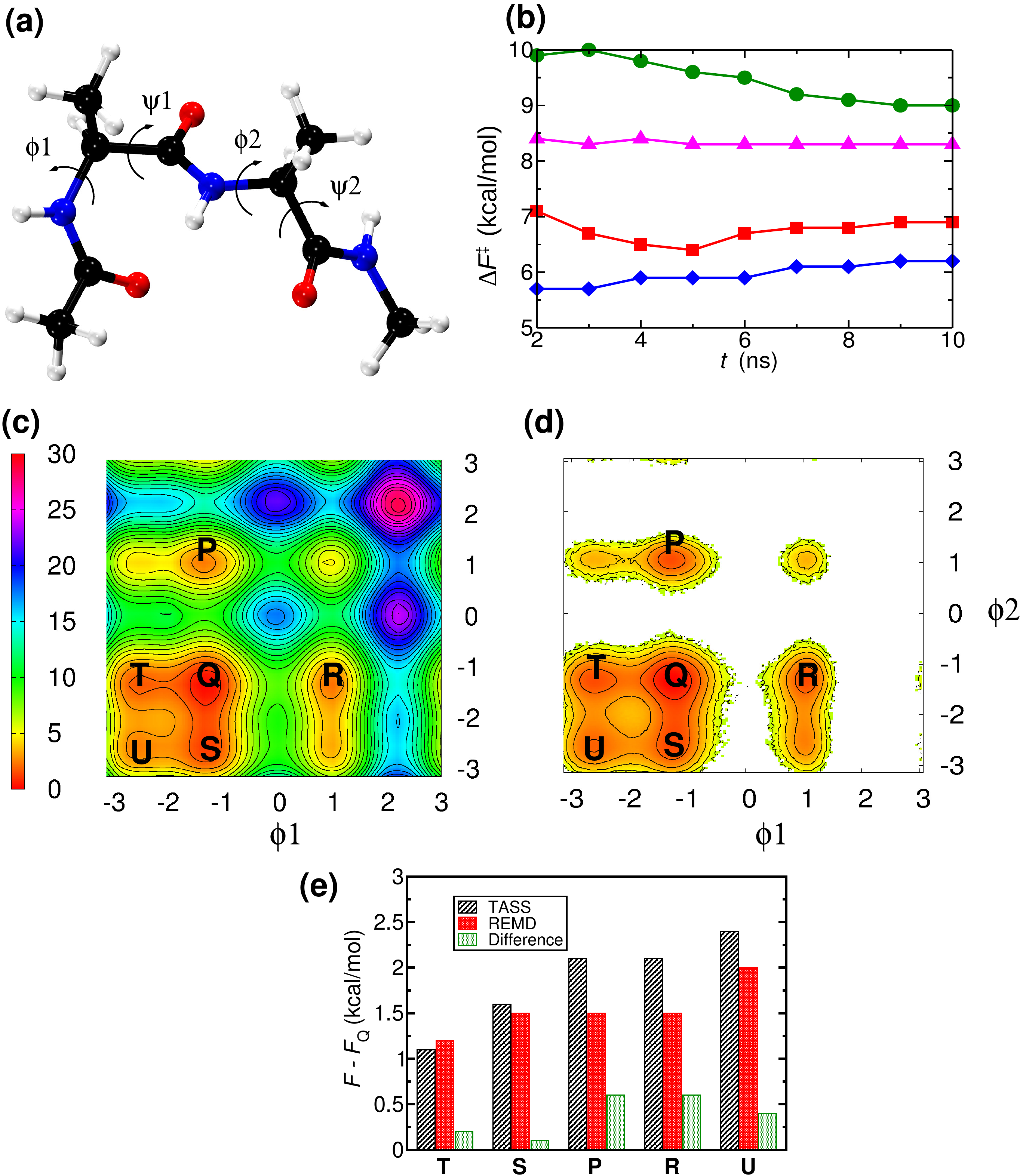}
\end{center}%
\caption[]{%
\label{alatrip_FES}%
{(a) Ball and stick representation of alanine tripeptide. $\phi$ and $\psi$ are defined as dihedral angle between atoms C--N--C$_{\alpha}$--C and N--C$_{\alpha}$--C--N, respectively as shown in the figure. 
color code: H (white), C (black), O (red), and N (blue).
(b) Convergence of free energy barriers as a function of simulation time per umbrella window.
Here symbols $\blacksquare$, $\bullet$, $\blacklozenge$, $\blacktriangle$ represent free energy barriers for
{\bf P}$\rightarrow${\bf Q}, {\bf Q}$\rightarrow${\bf P}, {\bf R}$\rightarrow${\bf Q}, and {\bf Q}$\rightarrow${\bf R}, respectively.
Projection of the reconstructed five--dimensional free energy surface on ($\phi1$, $\phi2$) plane as obtained from (c) TASS simulation, and (d) REMD simulation.
Contour values are shown for every $1$~kcal~mol$^{-1}$. Free energy is in kcal~mol$^{-1}$.
(e) Converged free energy of all the minima with respect to that of minimum {\bf Q} from TASS and REMD simulations are shown together with their difference.
}}
\end{figure*}

The free energy surface of alanine tripeptide (\fref{alatrip_FES}a) {\em in vacuo} 
as a function of four  backbone angles $(\phi1,\psi1,\phi2,\psi2)$
is explored here.
Alanine tripeptide was modeled using the ff14SB force--field~\cite{ff14SB} and MD simulations were carried out using the 
PLUMED--AMBER interface.~\cite{plumed,amber12} 
The time step was chosen as $1.0$~fs.
%
%We chose $\phi1$, $\phi2$, $\psi1$ and $\psi2$ as collective variables for sampling.
%

Here, umbrella bias was applied along the $\phi1$ while MTD bias was applied along the $\phi2$. 
All the four coordinates were sampled using high temperature.
MTD bias potentials were updated every $500$~fs and the parameters $w_0=0.6$~kcal~mol$^{-1}$, $\delta s= 0.05$~radians and 
$\Delta T = 900$~K were taken.  
Umbrella potentials were placed from $-\pi$ to $\pi$ at an interval of $0.2$~radians with $\kappa_{\rm h} = 1.2\times 10^2$~kcal~mol$^{-1}$~rad$^{-2}$,
$k_\alpha = 1.2 \times 10^3$~kcal~mol$^{-1}$~rad$^{-2}$, and a mass of $50$~a.m.u.~\textrm{\AA}$^2$~rad$^{-2}$ was assigned to all the auxiliary
variables. 

Initial structure for any given umbrella window was generated arbitrarily
 by setting the $\phi1$ internal coordinate to 
the equilibrium of the umbrella window, while the other
collective variables were corresponding to the minimum {\bf P}.
Langevin thermostat with a frictional coefficient of $0.001$~fs$^{-1}$
 was used for maintaining the temperature of physical system at $300$~K.
An overdamped Langevin thermostat with a friction coefficient of $0.1$~fs$^{-1}$ was used to maintain the CV temperature at $900$~K.
Before starting a TASS simulation, we carried out equilibration at $300$~K for a particular umbrella window for about $100$~ps.

For the purpose of comparison, we performed about $1$~$\mu s$ long replica exchange molecular dynamics (REMD) using AMBER~12.
Four replicas at temperatures $300$~K, $365$~K, $440$~K, and $535$~K were chosen.
Each replica was first equilibrated at its target temperature for $1$~ns.
An exchange attempt between replica was made at every $10$~ps.

The free energy surface along the ($\phi_1$, $\phi_2$) coordinates computed from the REMD simulations 
is given in \fref {alatrip_FES}d.
Six major minima were obtained, labeled as {\bf P}, {\bf Q}, {\bf R}, {\bf S}, {\bf T}, and {\bf U}. %TODO2:(DONE) make labeling also on Fig. 2d.
Subsequently, we carried out TASS simulation with four collective variables as mentioned before ($\phi1$, $\psi1$, $\phi2$, $\psi2$).
Computed free energy barriers separating these minima as a function of simulation time is plotted in \fref {alatrip_FES}b.
The barriers systematically converge, with an error less than 0.1~kcal~mol$^{-1}$, after 10~ps long simulation
per umbrella window.
The converged high dimensional surface is then projected to the ($\phi1$, $\phi2$) space; \fref{alatrip_FES}c.
Clearly, the positions of the minima and  the saddles are very well reproduced from TASS.
Moreover, the diagonal symmetry of the landscape can also be noticed, 
showing that the exploration of the high
dimensional free energy landscape has been performed very efficiently.
Similar observations were also made when the free energy surface was projected 
along the ($\phi1$, $\psi1$) and ($\phi2$, $\psi2$); see Figure~SI1. %TODO4 (SI or ESI??)
As free energy barriers could not be accurately computed from the REMD results (due to the insufficient sampling near the saddle points), we compare the free energy difference
between the minima obtained from REMD and TASS; see \fref {alatrip_FES}e and Table~SI3.
%\tref{t:tripep:conv}.
%
After convergence, the maximum difference between  the REMD and the TASS results is only $~0.6$~kcal~mol$^{-1}$, 
and this difference is likely due to the insufficient sampling in REMD.
These results further support that TASS can efficiently explore the high dimensional free energy landscapes
 and provide converged free energy estimates in a computationally efficient way.

\subsection {1,3--Butadiene to Cyclobutene Reaction}
\label{s:res:butadiene}

\begin{figure*}[]
\begin{center}%
\includegraphics[width=0.6\textwidth]{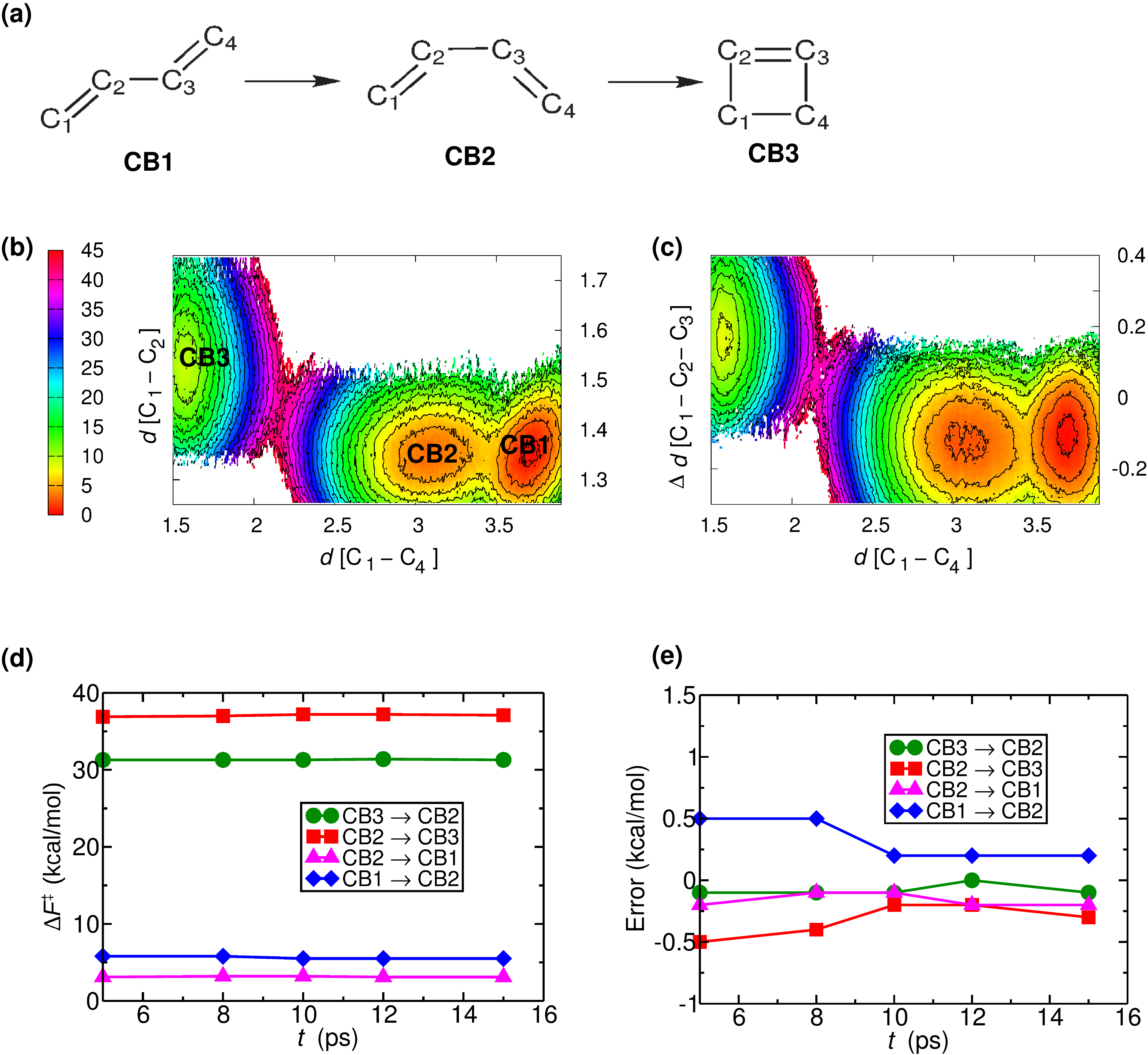}
\end{center}%
\caption[]{%
\label{buta_FES}%
(a) Structures of {\em trans}-1,3-butadiene ({\bf CB1}),{\em cis}-1,3-butadiene ({\bf CB2}), and cyclobutene ({\bf CB3});
(b) Projected free energy surface computed from TASS after 15~ps of the simulation per umbrella window; (c) Converged free energy surface computed from WS--MTD;
Free energy values are in kcal~mol$^{-1}$ relative to the free energy of the minimum ({\bf CB1});
Contour values are drawn between 1.0 and 45.0 kcal~mol$^{-1}$ at every 2~kcal~mol$^{-1}$ intervals; CVs are in \textrm{\AA};
(d)  Free energy barriers computed from TASS simulation as a function of simulation time per umbrella window; 
(e) Difference in the free energy barriers computed from TASS and WS--MTD (``Error'') as a function of simulation time per umbrella window. 
}
\end{figure*}

Here we explore the broad free energy surface for the conversion of 1,3--butadiene to cyclobutene which occurs via an electrocyclic reaction (see also \fref{buta_FES}a). 
%1,3--Butadiene itself can exist in {\em cis} and {\em trans}. 
%
%Free energy surface has a broad reactant basin and $3$~CVs can be used to observe this phenomenon.
%%
%Thus TASS method is used for performing this {\em ab initio} MD.
%%

We have chosen the following collective variables  to model this reaction: 
a) distance C$_1$--C$_4$, $d[\mathrm C_1-\mathrm C_4]$; 
b) the distance C$_1$--C$_2$, $d[\mathrm C_1-\mathrm C_2]$; 
c) the distance C$_2$--C$_3$, $d[\mathrm C_2-\mathrm C_3]$.

In TASS simulations, umbrella bias was applied along the $d$[C$_1$--C$_4$] coordinate and 
MTD bias was applied along $d[\mathrm C_1-\mathrm C_2]$.
Auxiliary variables corresponding to all the three coordinates were sampled using high temperature.
Simulations were carried out using 
{\em ab initio} MD employing plane--wave Kohn-Sham density functional theory (DFT) as
available in the CPMD program.~\cite{cpmd1}
PBE exchange correlation functional~\cite{PBEGGA} with ultrasoft pseudopotential~\cite{Pseudopotential} was used
here.
A cutoff of 30~Ry was used for the plane--wave expansion of wavefunctions.
System was taken in a cubic supercell of side length $15$~{\AA}.
Car--Parrinello~\cite{car85}
MD at $300$~K was carried out 
%using Nos$\mathrm{\grave{e}}$--Hoover chain thermostats.~\cite{nhc}
%%
with a time step of $0.096$~fs  and fictitious masses of orbitals were taken as $600$~a.u. 
%was assigned to the orbital degrees of freedom.
%

%
The parameter $k_\alpha$ was set to $1.2 \times 10^3$~kcal~mol$^{-1}$~\textrm{\AA}$^{-2}$ and $\mu_\alpha$ was $50.0$~a.m.u. 
Langevin thermostat with a friction coefficient of $0.4$~fs$^{-1}$ was used to maintain the temperature of the extended degrees of freedom to $600$~K.
In our simulations, $w_0=0.6$~kcal~mol$^{-1}$ and $\delta s=0.05$~Bohr were taken.
MTD bias was updated every $19$~fs. 
In US, the umbrella windows were placed from 1.5~\textrm{\AA} to 3.9~\textrm{\AA} at an interval of 0.05~\textrm{\AA} with $\kappa_{\rm h} = 4.4 \times 10^2$~kcal~mol$^{-1}$~\textrm{\AA}$^{-2}$.
Before starting the TASS simulation, each umbrella was equilibrated for about $2$~ps, and the initial structure
for each umbrella window was obtained arbitrarily, as done in the case of alanine tripeptide.

%
%Here we demonstrate that TASS simulation can be practically used in studying chemical reactions using AIMD simulations.
%%
%In particular, we study the cyclization of {\em trans} 1,3--butadiene ({\bf CB1})  (\fref{buta_FES}a) using three CVs.
%%
%A total of 49 umbrella windows were used, and 
We compare the results of the TASS simulation with the free energy surface and the
barriers computed using the WS-MTD approach from our earlier work.~\cite{shalini:2016} %TODO2 (DONE)
Free energy barriers converge to less than 0.5~kcal~mol$^{-1}$ (in comparison with  the WS-MTD barriers) within 10~ns per umbrella window;
see \fref{buta_FES}d,e.
Simulation for 5~ps seems enough to compute the free energy barriers with an error less than 0.5~kcal~mol$^{-1}$ (see Table~SI4 ).
The converged difference in the barriers of about 0.25~kcal~mol$^{-1}$  could be ascribed 
to the  differences in the type and  the number of collective variables used in TASS and WS-MTD.

These results show that the TASS approach could efficiently sample a high dimensional free energy landscape
 in three collective variable space of a chemical reaction.
The method seems to be as accurate as the WT-MTD, and is much efficient than the ordinary 
well-tempered MTD approach where free energy barriers for the same reaction was found not to 
converge even after $1000$~ps.~\cite{shalini:2016}

\subsection {Tetrahedral Intermediate Formation during 
Hydrolysis of Aztreonam and Class--C $\beta$--Lactamase complex}
\label{s:res:cbl}
\begin{figure*}[]
\begin{center}%
\includegraphics[width=0.7\textwidth]{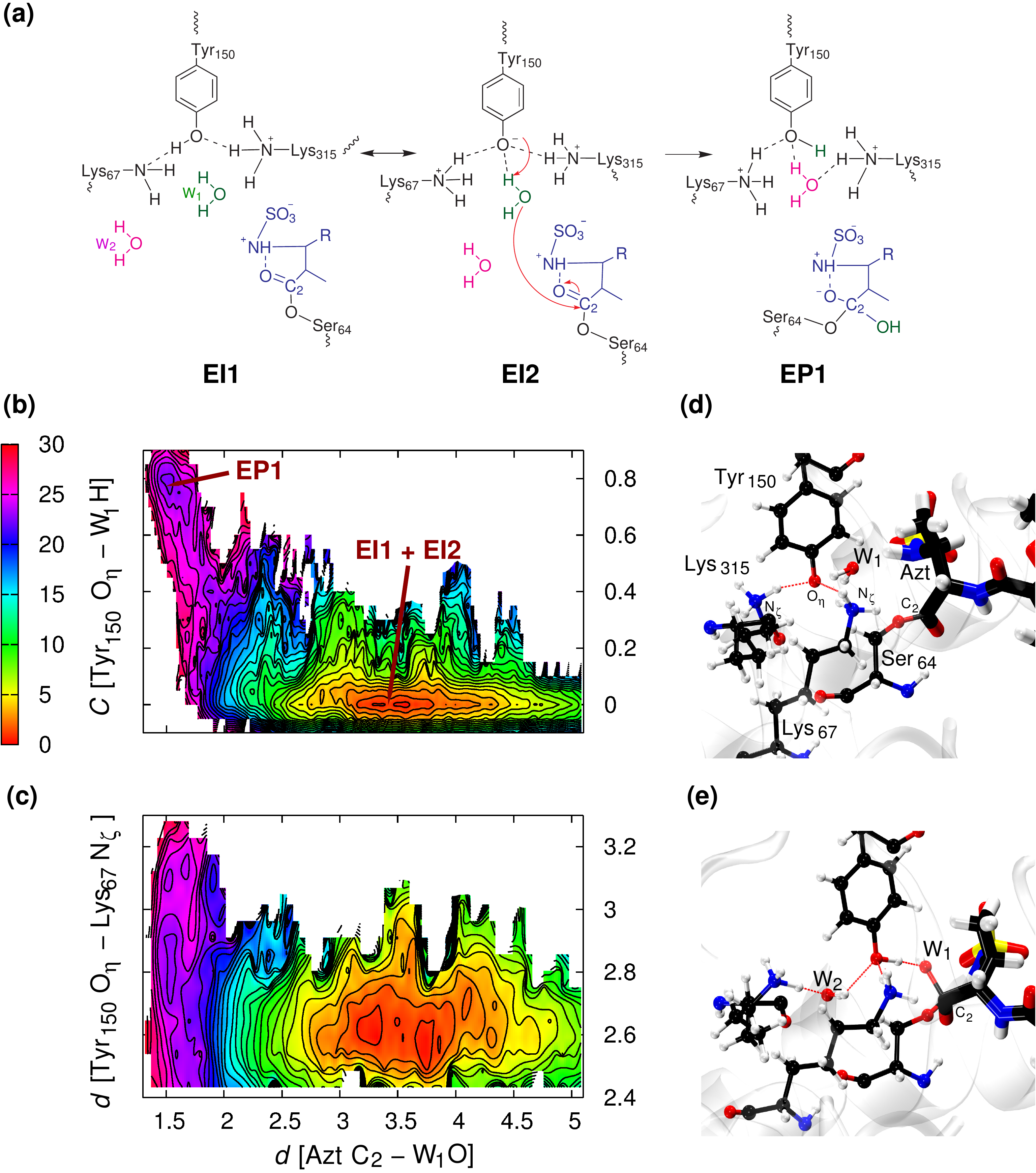} 
\end{center}%
\caption[]{%
\label{cbl_reaction}%
(a) Mechanism of formation of {\bf EP1} from the enzyme--drug covalent complex {\bf EI}
formed by aztreonam (blue color) and Class C $\beta$--lactamase. %TODO2 W1 and W2 should be labeled and W2 may be given magenta color (DONE)
Here W$_1$ molecule is activated by Tyr$_{150}$ and the former attacks C$_2$ resulting in {\bf EP1}; 
(b) and (c) are the two different projections of the five dimensional free energy landscape; (d) and (e) show
snapshots of {\bf EI2} and {\bf EP1} from the QM/MM trajectory; atom colors: S (yellow), O (red), N (blue), C (black), 
H (white) ; protein backbone is represented as transparent ribbons. 
}
\end{figure*}
To further demonstrate the application of the TASS method, we have applied this 
to model an enzymatic reaction
in a  DFT based QM/MM MD simulation.
Here we model the formation of a tetrahedral intermediate 
during the hydrolysis of the covalent complex formed by aztreonam drug and
Class--C $\beta$--Lactamase; see~\fref{cbl_reaction}. 
%
%We simulate this reaction using DFT based QM/MM MD techniques.
%
Four collective variables were chosen for simulating this hydrolysis reaction (see \fref{cbl_reaction} for  labeling): 
a) coordination number of ${\rm Tyr_{150}}$O$_{\eta}$ to hydrogens of W$_1$, 
$C[{{\rm Tyr}_{150}}{\rm O}_{\eta} - {\rm W_1 H}]$;  %TODO2 I changed CN to C
b) distance between ${\rm Azt}$C2 and ${\rm W_1 O}$, $d[\rm {\rm Azt C2} - {\rm W_1 O}]$; 
c) the distance ${\rm Tyr_{150}}$O$_{\eta}$ to Lys$_{67}$N$_\zeta$, 
$d[{\rm Tyr_{150} O}_\eta - {\rm Lys_{67} N}_\zeta]$;
d) the distance ${\rm Tyr_{150}}$O$_{\eta}$ to Lys$_{315}$N$_\zeta$, 
$d[{\rm Tyr_{150} O}_\eta - {\rm Lys_{315} N}_\zeta]$.
Here
\[  C[{{\rm Tyr}_{150}}{\rm O}_\eta - {\rm W_1 H}]   = \sum_{J \in \rm W_1 H }  \frac{ 1 }{ \left(1 + \left ( \frac{d_J}{d_0}   \right )^6 \right) } \]   %TODO2 (DONE)check
%\[  C[{{\rm Tyr}_{150}}{\rm O}_\eta - {\rm W_1 H}]   = \sum_{J \in \rm W_1 H }  \frac{ 1 }{ 1 + \left ( \frac{d_J}{d_0}   \right )^6 } \]   %TODO2 (DONE)check
where $d_J \equiv  d[{\rm Tyr}_{150}{\rm O}_\eta - {\mathrm{W_1 H}}]$ and $d_0=1.3$~{\AA}. %TODO2 check (DONE)
The auxiliary variables corresponding to all the four collective variables were sampled using high temperature (1000~K), while the
physical system was sampled at 300~K.
Here $C[{{\rm Tyr}_{150}}{\rm O}_{\eta} - {\rm W_1 H}]$ was chosen as a collective variable
to accelerate proton transfer from water to  Tyr$_{150}$O$_\eta$ 
and MTD bias was applied along this collective variable. 
To enhance the nucleophilic attack of OH$^-$ on the carbonyl carbon of the drug molecule 
$d[\rm {\rm Azt C2} - {\rm W_1 O}]$ coordinate was chosen as a collective variable which was sampled using
the US bias.
%
%For this CV, 37 umbrellas from a distance of 1.3~{\AA} to 5.1~{\AA} at intervals of 0.1~{\AA} were placed.
%
The collective variables $d[{\rm Tyr_{150} O}_\eta - {\rm Lys_{67} N}_\zeta]$ and 
$d[{\rm Tyr_{150} O}_\eta - {\rm Lys_{315} N}_\zeta]$ were considered for sampling 
different conformations of Tyr$_{150}$, Lys$_{67}$, and Lys$_{315}$.

The hybrid QM/MM simulations were performed using the CPMD/GROMOS interface~\cite{laio02a} 
as implemented in the CPMD package.
Aztreonam drug, side chains of Lys$_{67}$, Tyr$_{150}$, Ser$_{64}$, Lys$_{315}$, Thr$_{316}$; 
backbone of Lys$_{315}$, Thr$_{316}$, Gly$_{317}$, 
and two water molecules near the active site were treated quantum mechanically.
Rest of the protein and the solvent molecules were treated by molecular mechanics (MM).
%
%The initial structure of the enzyme-drug complex was taken from our earlier work.\cite{XX} %TODO2
%
The initial structure of the enzyme-drug complex was taken from the X--ray crystal structure 
corresponding to PDB ID 1FR6~\cite{1FR6}.
This whole system is composed of the enzyme, 
$11391$ TIP3P water molecules, %\cite{XX} %TODO2 
$2$ Na$^+$ ions, and $2$ Cl$^-$  ions,
%TODO2 Any anions added? Check- this has to be consistent with the JACS paper of Ravi
and was taken in a
periodic simulation box with the size 81.3$\times$75.6$\times$64.5~\textrm{\AA}$^{3}$.
Before starting the QM/MM simulation, MM MD simulation was carried out using the sander module 
in the AMBER suite of programs.\cite{amber12} %TODO4
The whole protein was treated using parm99 AMBER force--field~\cite{PARM99} whereas GAFF force--field~\cite{GAFF} was
employed for describing the drug molecule.
Restrained electrostatic potential charges for drug and Ser$_{64}$ complex were computed using the RED software.\cite{RED}  
During classical simulation, a time step of $1$~fs and a cut--off distance of $15$~\textrm{\AA} 
was used for non--bonded interaction.
After initial steps of minimization, $1$~ns of $NPT$ simulation was carried out 
using Langevin thermostat at $300$~K and Berendsen barostat at $1$~atm.
Subsequently, $10$~ns $NVT$ simulation was performed with the equilibrated density.
%
%The equilibrated system obtained after $NVT$ simulation was used for further QM/MM simulations.
%
%%%%%%%
In hybrid QM/MM simulations, QM part was treated using the plane--wave DFT with PBE exchange correlation functional.~\cite{PBEGGA}
Ultrasoft psuedopotentials~\cite{Pseudopotential} were chosen and  a plane--wave cutoff of 25~Ry was used.
A cubic QM box with a side length of 25.3~\textrm{\AA} was taken.
MM part of the system was treated using the param99~\cite{PARM99}
AMBER force--field.
Capping hydrogen atoms were added to saturate the bonds at the QM/MM boundary.
Capping hydrogen atoms were introduced between C$_\beta$ and C$_\gamma$ atoms of Tyr$_{150}$,
C$_\alpha$ and C$_\beta$ atoms of Ser$_{64}$,
C$_\gamma$ and C$_\delta$ atoms of Lys$_{67}$, and 
C$_\alpha$ and N atoms of Lys$_{315}$ and Gly$_{317}$. 

Constant temperature Car--Parrinello~\cite{car85}
MD at $300$~K was carried out using the 
Nos$\mathrm{\grave{e}}$--Hoover chain thermostats for the nuclei and orbital degrees of freedom.~\cite{nhc}
%TODO2 : are you sure you used 0.096 fs?DONE
A time step of $0.14$~fs was used to integrate the equations of motion and 
the fictitious masses of orbitals were taken as 600 a.u.
We assigned $k_\alpha = 1.2 \times 10^3$~kcal~mol$^{-1}$ and $\mu_\alpha = 50.0$~a.m.u.  
for all the auxiliary variables.
An overdamped Langevin thermostat with a friction coefficient of $0.4$~fs$^{-1}$ was 
used to maintain temperature of the auxiliary variables at $1000$~K.
In US, windows were placed from 1.3~\textrm{\AA} to 5.1~\textrm{\AA} at an interval of 
0.1~\textrm{\AA} with $\kappa_{\rm h} = 4.5 \times 10^2$~kcal~mol$^{-1}$~\textrm{\AA}$^{-2}$.
The MTD parameters were  $w_0=0.6$~kcal~mol$^{-1}$, $\delta s=0.05$~Bohr and $\Delta T= 2000$~K were taken.
MTD bias was updated every $19$~fs.

Before starting the TASS simulations, each umbrella window was equilibrated for about $4$~ps.
Initial structure for an umbrella window was taken from the adjacent equilibrated window. 
The whole protein, including the QM part, and the solvent molecules were free to move during the MD simulations.

%
%We carried out TASS simulation of the chemical reaction {\bf EI1}$\rightarrow${\bf EP1}
%using DFT based QM/MM MD protocols.
%%
%Four CVs were necessary to sample the activation of the attacking water molecule, nucleophilic
%attack and different conformations of Tyr$_{150}$ , Lys$_{67}$, and Lys$_{315}$ in the active site.
%%
%The importance of conformation sampling of Tyr$_{150}$, Lys$_{67}$, and Lys$_{315}$ was noted in an earlier work.~\cite{Ravi:2016} %DONE
%

%
We could successfully simulate the reaction {\bf EI1}$\rightarrow${\bf EP1} using the TASS method, and the converged reconstructed 
free energy surface  is given in \fref{cbl_reaction}b,c. %TODO2 (DONE)
Unlike in the previous benchmark cases, we have used varying simulation lengths (4-8~ps each) for different umbrella windows till a convergence in the free energy barrier was achieved. 
In the reactant basin, we have noticed proton transfer between Tyr$_{150}$ and Lys$_{67}$, i.e. 
{\bf EI1}$\leftrightarrow${\bf EI2}.
The hydrogen bonding interactions between the two residues were maintained throughout the reaction.
However, distance between the Tyr$_{150}$O$_\eta$ and Lys$_{67}$N$_\zeta$ increases
as proton transfer occurs from W$_1$ to Tyr$_{150}$O$_\eta$; see \fref{cbl_reaction}c.
On the other hand, the hydrogen bonding interaction between Lys$_{315}$ and Tyr$_{150}$ was
broken in the initial stages of the chemical reaction, as a result of which, another water molecule (W2)
moved into the active site, and Lys$_{315}$ formed interactions with Glu$_{272}$.

The free energy barrier for the reaction was computed from the projected free energy surface on 
$d[{\rm Tyr_{150} O}_\eta - {\rm Lys_{67} N}_\zeta]$ %TODO2 (DONE)
and 
$C[{{\rm Tyr}_{150}}{\rm O}_{\eta} - {\rm W_1 H}]$ %TODO2 (DONE) 
coordinates, and is 
24.5~kcal~mol$^{-1}$. %TODO2 (DONE)
From experimental studies~\cite{Monnaie} it is known that aztreonam is a slowly hydrolyzing drug 
and from the measured rate constants for deacylation, we estimate 
the corresponding free energy barrier as 23~kcal~mol$^{-1}$ (using the transition state theory).
This agrees well with our computed free energy barrier of 
24.5~kcal~mol$^{-1}$. %TODO2 (DONE)

The same reaction was failed to simulate in an ordinary MTD run, likely due to the broad nature of the basin 
along the $d[{\rm Tyr_{150} O}_\eta - {\rm Lys_{67} N}_\zeta]$ coordinate (see also \fref{cbl_reaction})c, %TODO2: (DONE)refer free energy surface 
and thus large computational time would be 
required to build the sufficient bias potential in the relevant parts of the free energy surface.
Moreover, W$_1$ water molecule was also driven out  of the active site, which 
was then replaced by an MM water molecule from the bulk (data not shown).
Both these difficulties are overcome in the TASS simulation since
US was carried out along the $d[{\rm Tyr_{150} O}_\eta - {\rm Lys_{67} N}_\zeta]$ coordinate.
Additionally, sampling of different conformations of the active site residues
%, especially the distance between Lys$_{67}$ 
%and Tyr$_{150}$, and the orientation of Lys$_{315}$ 
needed four collective variables
which is also practically difficult to sample properly with conventional MTD, especially in DFT based QM/MM simulations.

\section{Conclusions}
In this paper, we presented a method called TASS that combines
MTD, US, and TAMD to sample large number of collective variables 
and explore high dimensional free energy landscapes.
Free energy estimates using TASS is shown to converge systematically to the exact values.
We have demonstrated the efficiency of TASS  in sampling four and five dimensional free energy landscapes 
with multiple minima along different coordinates.
Moreover, the method is also shown to be practically usable for computing free energy surfaces of 
chemical reactions in {\em ab initio} and DFT based hybrid QM/MM MD simulations.
The method is well suited for exploring free energy surfaces that are broad and unbound, where
conventional enhanced sampling approaches such as MTD and TAMD become inefficient.
%
%The method can be viewed as an improvement of the conventional US method as a large number of 
%collective variables can be simultaneously sampled.
%%
Controlled exploration of free energy surfaces, for e.g. along certain reaction pathways, can also be achieved
in TASS, by an appropriate choice of the US coordinate.
The method permits one to add and remove collective variables in different umbrella windows, giving
flexibility and computational efficiency in exploring complex high dimensional free energy landscapes.
Although, we have used US and MTD biases, the method 
can be straightforwardly extended to the cases where MTD bias is not required (either in selected or for all the umbrella windows), by setting
$A(\tau)= 1$ in  \eref{e:p:unb}.
Replica exchange based algorithms can also be
combined with TASS to further improve the sampling efficiency.
The TASS Hamiltonian in \eref{eqn:ham} can be realized effortlessly in simulations using  MD plugins 
like PLUMED~\cite{plumed} which has been interfaced with several popular MM and QM programs.

\begin{acknowledgments}
Authors thank IIT Kanpur for availing the HPC facility.  SA thanks UGC for Ph.~D fellowship.
\end{acknowledgments}
%

%\section*{Supplementary Information}
%See supplementary information

%\bibliographystyle{aipnum4-1.bst}
%\bibliography{myrefs_thesis}

\end{document}